\documentclass[a4paper,10pt,notitlepage,twocolumn]{revtex4-1}
\usepackage[utf8]{inputenc}
\usepackage[ngerman,english]{babel}
\pdfoutput=1
\usepackage{url}
\usepackage{bbm}
\usepackage[T1]{fontenc}
\usepackage[utf8]{inputenc}
\usepackage{graphicx}
\usepackage{eurosym}
\usepackage{amsmath}
\usepackage{natbib}
\usepackage{amssymb}
\usepackage{hyperref}
\usepackage{amsthm}
\usepackage{subfigure}
\usepackage{setspace}
\usepackage[titletoc,title]{appendix}
\usepackage[left=2.0cm,right=2.0cm,top=3.0cm,bottom=3.0cm]{geometry}
\newcommand\numberthis{\addtocounter{equation}{1}\tag{\theequation}}
\newtheorem{Theorem}{Theorem}

\begin{document}
\title{Quantum error correction against photon loss using NOON states}
\author{Marcel Bergmann}
\author{Peter van Loock}
\affiliation{Institut für Physik, Johannes Gutenberg-Universität, Staudingerweg 7, 55128 Mainz, Germany}
\date{\today}
\begin{abstract}
\noindent
The so-called NOON states are quantum optical resources known to be useful especially for quantum lithography and metrology. At the same time, they are known to be very sensitive to photon losses and rather hard to produce experimentally.
Concerning the former, here we present a scheme where NOON states are the elementary resources for building quantum error correction codes against photon losses, thus demonstrating that such resources can also be useful
to suppress the effect of loss. Our NOON-code is an exact code that can be systematically extended from one-photon to higher-number losses. Its loss scaling depending on the codeword photon number is the same as for
some existing, exact loss codes such as bosonic and quantum parity codes, but its codeword mode number is intermediate between that of the other codes. Another generalisation of the NOON-code is given for arbitrary logical qudits
instead of logical qubits. While, in general, the final codewords are always obtainable from multi-mode NOON states through application of beam splitters, both codewords for the one-photon-loss qubit NOON-code can be simply created from single-photon
states with beam splitters. We give various examples and also discuss a potential application of our qudit code for quantum communication.
\end{abstract}
\maketitle

\section{Introduction}
\noindent
So-called NOON states are an important resource in optical quantum information science \cite{Sanders,Downling1, Mitchell, Walther}. 
They are bipartite entangled, $N$-photon two-mode states where the $N$ photons occupy either one of two optical modes, $\frac{1}{\sqrt{2}}(|N0\rangle+|0N\rangle)$.
NOON states have been widely used in quantum communication \cite{Gisin}, quantum metrology \cite{metrology} and quantum lithography \cite{lithography},
because they allow for super-sensitive measurements, e.g. in optical interferometry. This is related to the substandard quantum-limit behaviour of NOON states, i.e.  a factor $\sqrt{N}$ improvement to the shot noise limit can be achieved \cite{Rohde}.
Due to their practical relevance, various schemes for NOON state generation based on strong non-linearities \cite{Gerry, Kapale} 
or measurement and feed-forward \cite{Cerf, VanMeter, Cable} have been proposed. Unfortunately, NOON states are very fragile, which focused 
recent research on their entanglement and phase properties in noisy environments \cite{Sperling} or on the enhancements of NOON state sensitivity by non-Gaussian operations \cite{Jeffers}.\\
Though very sensitive to losses, NOON states can be useful resources to build quantum error correcting codes, as will be shown in this paper. Optical quantum information and especially their use in long-distance quantum communication suffers from loss. 
Here, the main mechanism of decoherence is photon loss which is theoretically described by the amplitude-damping (AD) channel that acts on each field mode. To protect quantum information from photon loss, various kinds of quantum error correction codes for AD were proposed \cite{Banaszek, bosonic_codes, Ralph}. 
In this context, it was also observed that quantum error correction codes fall in one of two classes: exact or approximate. 
The usual quantum error correction conditions are strictly fulfilled by exact quantum codes, whereas approximate codes only fulfil a set of relaxed conditions \cite{approximate_codes}. \\
Recently, quantum parity codes (QPCs) \cite{lutkenhaus, Ralph} as an example of exact codes for AD were employed in the  context of long-distance quantum communication. 
A QPC consumes $N^{2}$ photons distributed in $2N^{2}$ modes with at most one photon per mode. This code therefore requires a maximal number of modes. In contrast, an important class of exact AD codes, the so-called bosonic codes
using $N^{2}$ total photons,  introduced in \cite{bosonic_codes}, may use no more than just two modes  at the expense of having up to $N^{2}$ photons per mode. The codes to be developed
in this paper are intermediate between QPC and bosonic codes, because they use $N^{2}$ photons in  $2N$ modes with at most $N$ photons per mode. Our code is a block code 
like QPC and unlike the general bosonic code, with the same number of blocks as QPC, but with the $N^{2}$ photons distributed among a smaller number of modes in every block compared to QPC. Despite their structural differences,
all these loss codes, including our code,  protect a logical qubit exactly against $N-1$ photon losses using $N^{2}$ photons. Thus, only in our scheme, both the total mode number and the maximal photon number per mode scale linearly with $N$ to achieve protection against $N-1$ losses.
Another crucial difference, compared to QPC and bosonic codes lies in the systematic accessibility of our codewords from NOON states and linear optics. 
For the simplest special case of a one-photon-loss qubit code, even only one-photon Fock states are sufficient as resources for codeword generation, as the $N=2$ NOON
state corresponds to the well-known Hong-Ou-Mandel state. Our systematic approach can be also applied to qudit-code constructions, while certain examples of bosonic qudit codes were also given in \cite{bosonic_codes} (see also the recent work in \cite{Grassl}).\\

\noindent The structure of the paper is as follows: in the second section, the AD model and the basics of quantum error correction are introduced and some known photon loss codes are 
reviewed. In the succeeding section, quantum codes for logical qubits are systematically developed.  
The third section discusses the extension of  this systematic scheme to logical qudits in a natural manner by switching from beam splitters to general $N$-port devices. 
It is shown that the scaling of the fidelity only depends on the total photon number and, especially, that it is independent of the dimension of the logical qudit. Section  5 presents an in-principle method for the generation of  an arbitrary logical qubit 
state for the one-photon-loss qubit code based on linear optics and light-matter interactions. The last section, as an example of an application, describes the use of the qudit code in a one-way quantum communication scheme. This scheme
does not intrinsically provide an optimal rate between physical versus logical qubits like another recent approach \cite{Jiang2}, but nonetheless allows 
for sending more quantum information at each time step with the same loss protection.

\section{Quantum error correction and photon loss}
\noindent Photon loss can be modelled by the AD channel. The non-unitary error operators $A_{k}$, specifying the loss of $k$ photons in a single mode, are given by \cite{bosonic_codes}
\begin{equation}
A_{k}=\sum\limits_{n=k}^{\infty}\sqrt{\binom{n}{k}}\sqrt{\gamma}^{n-k}\sqrt{1-\gamma}^{k}|n-k\rangle\langle n|,
\end{equation}
$\forall k\in \{0,1,\cdots,\infty\}$. Here, $\gamma$ is the damping parameter and $1-\gamma$ corresponds to the probability of losing one photon. The operators $A_{k}$ form a POVM, i.e. $A_{k}\geq 0$ and $\sum\limits_{k=0}^{\infty}A_{k}^{\dagger}A_{k}=1$. 
The non-unitary evolution of an arbitrary single-mode density operator $\rho$ under the effect of 
AD is 
\begin{equation}\label{AD}
\rho\rightarrow\rho_{f}=\sum\limits_{k=0}^{\infty}A_{k}\rho A_{k}^{\dagger}.
\end{equation}
By employing a quantum code, one is partially able to reverse the dynamics implied by Eq. \eqref{AD} and recover the original state. A proper quantum code enables one to detect and correct a certain set of errors on the encoded state. A quantum code is a vector space
spanned by basis codewords, denoted by $|\bar{0}\rangle\equiv |c_{1}\rangle$ and $|\bar{1}\rangle\equiv |c_{2}\rangle$ for a qubit code, and a subspace of some higher-dimensional Hilbert space. Normalized elements of this vector space of the form 
$\alpha|\bar{0}\rangle+\beta|\bar{1}\rangle$ are called logical qubits. This notion can be extended to qudit codes, where there are more than two codewords $|c_{1}\rangle,\cdots,|c_{d}\rangle$ to encode a logical $d$-level system.   
To form a proper quantum code, the logical basis codewords have to fulfil certain conditions. We state the famous Knill-Laflamme conditions which are a set of necessary and sufficient condition for the existence of a recovery operation \cite{N-C}:
\begin{Theorem}{(Knill-Laflamme)}\\
 Let $C=span\{|c_{1}\rangle,|c_{2}\rangle,\cdots,|c_{d}\rangle\}$ be a quantum code, $P$ be the projector onto $C$ and $\{E_{i}\}$ the set of error operators. There exists an error-correction operation $\mathcal{R}$ that corrects
 the errors $\{E_{i}\}$ on C, iff
 \begin{equation}\label{knill-laflamme}
 PE_{i}^{\dagger}E_{j}P=\Lambda_{ij}P,~~~ \forall i,j
 \end{equation}
 for some semi-positive, Hermitian matrix $\Lambda$ with matrix elements $\Lambda_{ij}$.
\end{Theorem}

\noindent For photon loss codes (in particular, those exact codes with a fixed total photon number), the matrix $\Lambda$ is typically diagonal, i.e. $\Lambda_{ij}=g_{i}\delta_{ij}$. This defines non-degenerate codes with different
loss errors (especially different numbers of photons lost, but also different modes subject to loss) corresponding to orthogonal error spaces. Nonetheless, certain instances of our code do exhibit
degeneracy for a given number of lost photons.\\
The Knill-Laflamme (KL) conditions contain two basic notions. The first notion is the orthogonality of corrupted codewords, i.e.
\begin{equation}\label{ortho}
 \langle c_{k}|E_{i}^{\dagger}E_{j}|c_{l}\rangle =0~~ \text{if}~~ k\neq l.
\end{equation}
The second one is the non-deformability condition, i.e.
\begin{equation}\label{deform}
  \langle c_{l}|E_{i}^{\dagger}E_{i}|c_{l}\rangle =g_{i},~~ \forall l.
\end{equation}
This means that the norm of a corrupted codeword only depends on the error operator and not on the codeword itself.\\
Before proceeding with the code construction, we highlight some examples of existing photon loss codes. 
In the first example, a logical qubit is encoded in a certain two-dimensional subspace of two bosonic modes. The basis codewords are chosen 
in the following way \cite{bosonic_codes}:
\begin{equation}\label{eq:Leung}
\begin{aligned}
|\bar{0}\rangle=&\frac{1}{\sqrt{2}}(|40\rangle+|04\rangle),\\
|\bar{1}\rangle=&|22\rangle,
\end{aligned}
\end{equation}
i.e. any logical qubit has a total photon number $N^{2}=4$. This code corrects exactly $N-1=1$ photon losses. The worst-case fidelity, as defined further below, is found to be 
$F=\gamma^{4}+4\gamma^{3}(1-\gamma)=1-6(1-\gamma)^{2}+8(1-\gamma)^{3}-3(1-\gamma)^{4}$.
In the same reference \cite{bosonic_codes}, the following code is given:
\begin{equation}
\begin{aligned}
|\bar{0}\rangle&=\frac{1}{\sqrt{2}}(|70\rangle+|16\rangle),\\
|\bar{1}\rangle&=\frac{1}{\sqrt{2}}(|52\rangle+|34\rangle).
\end{aligned}
\end{equation}
This code corrects also all one-photon losses and its worst-case fidelity is $\gamma^{7}+7\gamma^{6}(1-\gamma)=
1-21(1-\gamma)^{2}+70(1-\gamma)^{3}-105(1-\gamma)^{4}\linebreak +84(1-\gamma)^{5}-35(1-\gamma)^{6}+6(1-\gamma)^{7}$. 
\\Another example that encodes a qubit in three optical modes with a total photon number of 3 was 
proposed in \cite{Banaszek}. The basis codewords are 

\begin{equation}
\begin{aligned}
|\bar{0}\rangle&=\frac{1}{\sqrt{3}}(|300\rangle+|030\rangle+|003\rangle),\\
|\bar{1}\rangle&=|111\rangle.\\
\end{aligned}
\end{equation}
The fidelity in this case is $\gamma^{3}+3\gamma^{2}(1-\gamma)=1-3(1-\gamma)^{2}+2(1-\gamma)^{3}$. Moreover, note that all three codes 
given above are capable of exactly correcting only the loss of one photon, as can be easily seen by checking the KL conditions. An example 
for a proper two-photon-loss code is \cite{bosonic_codes}

\begin{equation}
\begin{aligned}
|\bar{0}\rangle&=\frac{1}{2}|90\rangle+\frac{\sqrt{3}}{2}|36\rangle,\\
|\bar{1}\rangle&=\frac{1}{2}|09\rangle+\frac{\sqrt{3}}{2}|63\rangle,\\
\end{aligned}
\label{eq: nine}
\end{equation}
whose worst-case fidelity is found to be $F=\gamma^{9}+9\gamma^{8}(1-\gamma)+36\gamma^{7}(1-\gamma)^{2}\approx 1-84(1-\gamma)^{3}$.\\
What these codes also have in common is their small number of optical modes, at the expense of having rather large maximal photon numbers in each mode 
(in order to obtain a sufficiently large Hilbert space). Conversely, a code that has at most one photon in any mode, but a correspondingly large total mode number,
is the QPC. The simplest non-trivial QPC, denoted as QPC(2,2), reads as follows \cite{Ralph}:
\begin{equation}
\begin{aligned}
 |\bar{0}\rangle&=\frac{1}{\sqrt{2}}(|10101010\rangle+|01010101\rangle),\\
 |\bar{1}\rangle&=\frac{1}{\sqrt{2}}(|10100101\rangle+|01011010\rangle).\\
\end{aligned}
\end{equation}
It also corrects exactly the loss of one photon. Different from all these codes that all consist of superpositions of states
with a fixed photon number is the following code \cite{approximate_codes}:

\begin{equation}
\begin{aligned}
 |\bar{0}\rangle=\frac{1}{\sqrt{2}}(|0000\rangle+|1111\rangle),\\
 |\bar{1}\rangle=\frac{1}{\sqrt{2}}(|0011\rangle+|1100\rangle).\\
\end{aligned}
\end{equation}
This code is conceptually distinct, because it does not satisfy the usual KL conditions. It satisfies certain relaxed conditions, 
which leads, in a more general setting, to approximate quantum error correcting schemes \cite{approximate_codes}. 
The above approximate code still satisfies the KL conditions up to linear order in $1-\gamma$, corresponding to one-photon-loss correction, while it requires
4 physical qubits (single-rail qubits encoded as vacuum $|0\rangle$ and single-photon $|1\rangle$) instead of 5 physical qubits for the minimal universal 
one-qubit-error code. Note that for dual-rail physical qubits (i.e., the approximate Leung code \cite{approximate_codes} concatenated with standard optical dual-rail encoding), one 
obtains QPC(2,2), which is then an exact one-photon-loss code.\\
After setting the stage, we will now start to discuss how to construct new quantum codes for AD to suppress the effect of photon losses.
\section{Qubit codes}
\label{sec: Qubit codes}
\noindent Let us consider the following qubit codewords defined in the three-dimensional Hilbert space of two photons distributed among
two modes,

\begin{equation}
\begin{aligned}
 |\bar{0}\rangle&=\frac{1}{\sqrt{2}}(|20\rangle+|02\rangle),\\
 |\bar{1}\rangle&=|11\rangle.\\
\end{aligned}
\end{equation}
The action of the AD channels on the two modes of the logical qubit $|\bar{\Psi}\rangle=c_{0}|\bar{0}\rangle+c_{1} |\bar{1}\rangle$ is
\footnote{In the remainder of the paper $c_{0},c_{1},..,c_{d-1}\in \mathbb{C}$ are the coefficients of our logical qudits.}
\begin{equation}
\begin{aligned}
 A_{0}\otimes A_{0}|\bar{\Psi}\rangle&=\sqrt{\gamma^{2}}|\bar{\Psi}\rangle,\\
 A_{1}\otimes A_{0}|\bar{\Psi}\rangle&=\sqrt{\gamma(1-\gamma)}(c_{0} |10\rangle+c_{1}|01\rangle)),\\
 A_{0}\otimes A_{1}|\bar{\Psi}\rangle&=\sqrt{\gamma(1-\gamma)}(c_{0} |01\rangle+c_{1}|10\rangle)),\\
\end{aligned}
\end{equation}
including the first three error operators $E_{1}=A_{0}\otimes A_{0}$, $E_{2}=A_{1}\otimes A_{0}$ and $E_{3}=A_{0}\otimes A_{1}$, of which the last
two describe the loss of a photon. Obviously, the one-photon-loss spaces are not orthogonal (they are even identical) and the qubit is subject to a random bit flip 
for the one-photon-loss case. A different choice would be: 
\begin{equation}
\begin{aligned}
|\bar{0}\rangle&=\frac{1}{2}|20\rangle+\frac{1}{2}|02\rangle+\frac{1}{\sqrt{2}}|11\rangle,\\
 |\bar{1}\rangle&=\frac{1}{2}|20\rangle+\frac{1}{2}|02\rangle-\frac{1}{\sqrt{2}}|11\rangle.
\end{aligned}
\label{eq: fake}
\end{equation}
After AD, this becomes:
\begin{align*}
 A_{0}\otimes A_{0}|\bar{\Psi}\rangle&=\sqrt{\gamma^{2}}|\bar{\Psi}\rangle,\numberthis \\
 A_{1}\otimes A_{0}|\bar{\Psi}\rangle&=\sqrt{\gamma(1-\gamma)}\\
&\times(c_{0} \frac{1}{\sqrt{2}}(|10\rangle+|01\rangle)+c_{1}\frac{1}{\sqrt{2}}(|10\rangle-|01\rangle)),\\
A_{0}\otimes A_{1}|\bar{\Psi}\rangle&=\sqrt{\gamma(1-\gamma)}\\
 &\times(c_{0} \frac{1}{\sqrt{2}}(|10\rangle+|01\rangle)-c_{1}\frac{1}{\sqrt{2}}(|10\rangle-|01\rangle)).
\end{align*}
\noindent
Here, the phase flip in the last line corresponds to a violation of the KL criteria,$\langle \bar{0}|E_{2}^{\dagger}E_{3}|\bar{0}\rangle\neq \langle\bar{1}|E_{2}^{\dagger}E_{3}|\bar{1}\rangle$,
preventing the encoding from being a proper quantum error correcting code. Indeed, again we have identical one-photon-loss spaces. One can easily verify that any choice of codewords 
will either lead to overlapping one-photon-loss spaces or the qubit is completely lost. A possible remedy is to construct codes composed of blocks.\\
To demonstrate this, we first deal with the specific example for encoding a logical qubit. Define 
\begin{equation}
\begin{aligned}
|t_{0}^{2,2}\rangle&=BS[|20\rangle]=\frac{1}{2}|20\rangle+\frac{1}{2}|02\rangle+\frac{1}{\sqrt{2}}|11\rangle,\\
|t_{1}^{2,2}\rangle&=BS[|02\rangle]=\frac{1}{2}|20\rangle+\frac{1}{2}|02\rangle-\frac{1}{\sqrt{2}}|11\rangle,
\end{aligned}
\end{equation}
as the "input states" for our encoding, where $BS[~]$ denotes a 50:50 beam splitter transformation.
Note that, in general, a beam splitter with reflectivity $r$ and transmittance $t$ acts on a two-mode Fock state as  

\begin{align*}
 |m,n\rangle\mapsto \sum\limits_{j,k=0}^{m,n}\sqrt{\frac{(j+k)!(m+n-j-k)!}{m!n!}}\binom{m}{j}\binom{n}{k}\\
 \times(-1)^{k}t^{n+j-k}r^{m-j+k}|m+n-j-k,j+k\rangle.\numberthis
\end{align*}

\noindent
In the case of a 50:50 beam splitter ($t=r=\frac{1}{\sqrt{2}}$), this reduces to
\begin{equation}
\begin{aligned}
 |m,n\rangle&\mapsto\sum\limits_{j,k=0}^{m,n}\sqrt{\frac{1}{2}}^{n+m}\sqrt{\frac{(j+k)!(m+n-j-k)!}{m!n!}}\binom{m}{j}\\
 &\times\binom{n}{k}(-1)^{k}|m+n-j-k,j+k\rangle,
\end{aligned}
\end{equation}

\noindent
and we obtain in particular:
\begin{equation}
\begin{aligned}
BS[|N0\rangle]&=\sqrt{\frac{1}{2}}^{N}\sum\limits_{j=0}^{N}\sqrt{\binom{N}{j}}|N-j,j\rangle,\\
BS[|0N\rangle]&=\sqrt{\frac{1}{2}}^{N}\sum\limits_{j=0}^{N}(-1)^{j}\sqrt{\binom{N}{j}}|N-j,j\rangle.
\end{aligned}
\end{equation}

\noindent
Now by means of a Hadamard-type operation on $|t_{0}^{2,2}\rangle$ and $|t_{1}^{2,2}\rangle$, the following states are obtained:

\begin{equation}
\begin{aligned}
|\widetilde{0}\rangle&=\frac{1}{\sqrt{2}}(|t_{0}^{2,2}\rangle+|t_{1}^{2,2}\rangle)=\frac{1}{\sqrt{2}}(|20\rangle+|02\rangle),\\
|\widetilde{1}\rangle&=\frac{1}{\sqrt{2}}(|t_{0}^{2,2}\rangle-|t_{1}^{2,2}\rangle)=|11\rangle.
\end{aligned}
\end{equation}
Note that $|\widetilde{1}\rangle$ equals $BS[\frac{1}{\sqrt{2}}(|20\rangle-|02\rangle)]$, whereas $|\widetilde{0}\rangle$ is the two-photon NOON state which is invariant
under the beam splitter transformation. A logical qubit can now be encoded according to
\begin{equation}\label{twophotons}
 |\bar{\Psi}\rangle=c_{0}|\widetilde{0}\rangle|\widetilde{0}\rangle+c_{1}|\widetilde{1}\rangle|\widetilde{1}\rangle\equiv c_{0}|\bar{0}\rangle+c_{1}|\bar{1}\rangle.
\end{equation}

\noindent
We prove in the following that the codewords
\begin{equation}
\begin{aligned}
|\bar{0}\rangle&=|\widetilde{0}\rangle|\widetilde{0}\rangle=\frac{1}{\sqrt{2}}(|20\rangle+|02\rangle)\frac{1}{\sqrt{2}}(|20\rangle+|02\rangle)\\
&=\frac{1}{2}(|2020\rangle+|2002\rangle+|0220\rangle+|0202\rangle),\\
|\bar{1}\rangle&=|\widetilde{1}\rangle|\widetilde{1}\rangle=|1111\rangle,
\end{aligned}
\end{equation}
form a quantum error correcting code for the AD channel. 
Calculating the action of AD on the basis codewords and checking the KL conditions, we obtain

\begin{align*}
 &A_{0}\otimes A_{0}\otimes A_{0}\otimes A_{0}|\bar{\Psi}\rangle\\
 &=\sqrt{\gamma^{4}}|\bar{\Psi}\rangle,\\
 &A_{1}\otimes A_{0}\otimes A_{0}\otimes A_{0}|\bar{\Psi}\rangle\\
 &=\sqrt{\gamma^{3}(1-\gamma)}\left(\frac{c_{0}}{\sqrt{2}}(|1020\rangle+|1002\rangle)+c_{1}|0111\rangle\right),\\
 &A_{0}\otimes A_{1}\otimes A_{0}\otimes A_{0}|\bar{\Psi}\rangle\\
 &=\sqrt{\gamma^{3}(1-\gamma)}\left(\frac{c_{0}}{\sqrt{2}}(|0120\rangle+|0102\rangle)+c_{1}|1011\rangle\right),\numberthis \\
 &A_{0}\otimes A_{0}\otimes A_{1}\otimes A_{0}|\bar{\Psi}\rangle\\
 &=\sqrt{\gamma^{3}(1-\gamma)}\left(\frac{c_{0}}{\sqrt{2}}(|2010\rangle+|0210\rangle)+c_{1}|1101\rangle\right),\\
 &A_{0}\otimes A_{0}\otimes A_{0}\otimes A_{1}|\bar{\Psi}\rangle\\
 &=\sqrt{\gamma^{3}(1-\gamma)}\left(\frac{c_{0}}{\sqrt{2}}(|2001\rangle+|0201\rangle)+c_{1}|1110\rangle\right).\\
\end{align*}

\noindent
The KL conditions are obviously fulfilled for one-photon-loss errors. Note that, after losing any two or more photons, the logical qubit cannot be recovered
anymore.\\
To be able to actively perform quantum error correction, it is a necessary task to determine the syndrome information, i.e. in our
case the location (the mode) where a photon loss occurred. To get this information, we first measure the total photon number per block. If
the result is "2" on each block, there is no photon missing and the logical qubit is unaffected. However, if for example a photon got lost on the first
mode, the result is "1" for the first block and "2" for the other. This result is not unique, because there are still two possible corrupted states with this 
measurement pattern. In order to resolve this, inter-block photon number parity measurements with respect to modes 2+3 and 1+4 are suitable. The results
"even-odd" and "odd-even" uniquely determine the corrupted state which can then be accordingly recovered. 
Note that all the measurements discussed here are assumed to be of QND-type such that also higher photon losses can be non-destructively detected. 
But so far these cannot be corrected by means of the encoding.\\
A convenient measure for the quality of a quantum error correcting code is the worst-case fidelity, defined as \cite{N-C,bosonic_codes}
\begin{equation}
 F=\min\limits_{|\bar{\Psi}\rangle\in C} \langle\bar{\Psi}|\mathcal{R}(\bar{\rho}_{f})|\bar{\Psi}\rangle,
 \label{eq: fidelity}
\end{equation}
where $\bar{\rho}_{f}$ is the  final mixed state after multi-mode amplitude damping (with the only assumption that each AD channel
acts independently on each mode) and $\mathcal{R}$ is the recovery operation. Note that the recovery operation
always exists if the KL conditions are fulfilled. The fidelity defined in Eq.\eqref{eq: fidelity} is a suitable figure of merit to assess
the performance of a quantum error correction code \footnote{The exact loss codes considered in this paper have identical worst-case and average fidelities.}. In particular, it also reveals if an encoding is not a proper code (see, e.g. the encoding
in Eq.\eqref{eq: fake}). In our case, the worst-case fidelity   
is easily calculated as
\begin{equation}
F=\gamma^{4}+4\gamma^{3}(1-\gamma)\approx 1-6(1-\gamma)^{2}.
\end{equation}
Note that this code has the same scaling as the four-photon-code of \cite{bosonic_codes} described by Eq.\eqref{eq:Leung}.\\
For higher losses, we can use NOON states with higher photon number to encode a logical qubit. For this purpose, let us define the input states for the codewords 
as 
\begin{equation}
\begin{aligned}
|t_{0}^{2,3}\rangle=BS[|30\rangle]&=\frac{1}{2\sqrt{2}}|03\rangle+\frac{1}{2}\sqrt{\frac{3}{2}}|12\rangle+\frac{1}{2}\sqrt{\frac{3}{2}}|21\rangle\\
&+\frac{1}{2\sqrt{2}}|30\rangle,\\
|t_{1}^{2,3}\rangle=BS[|03\rangle]&=-\frac{1}{2\sqrt{2}}|03\rangle+\frac{1}{2}\sqrt{\frac{3}{2}}|12\rangle-\frac{1}{2}\sqrt{\frac{3}{2}}|21\rangle\\
&+\frac{1}{2\sqrt{2}}|30\rangle,
\end{aligned}
\end{equation}
such that this time
\begin{equation}
\begin{aligned}
|\widetilde{0}\rangle&=\frac{1}{\sqrt{2}}(|t_{0}^{2,3}\rangle+|t_{1}^{2,3}\rangle)=\frac{1}{2}|30\rangle+\frac{\sqrt{3}}{2}|12\rangle,\\
|\widetilde{1}\rangle&=\frac{1}{\sqrt{2}}(|t_{0}^{2,3}\rangle-|t_{1}^{2,3}\rangle)=\frac{1}{2}|03\rangle+\frac{\sqrt{3}}{2}|21\rangle,
\end{aligned}
\end{equation}
\noindent
become the states after the Hadamard-type gate. We could now again build a qubit like in Eq.\eqref{twophotons}.
However, we find that the resulting six-photon two-block (four-mode) code only corrects certain two-photon losses and therefore there is no significant enhancement compared to the $N=2$ code above. This can be 
understood by looking at the corrupted logical qubit for losses of up to two photons. The details for this are presented in Appendix \ref{sec: inefficiency}.
The conclusion is that some of the orthogonality requirements are violated for certain two-photon losses which consequently cannot be corrected. 
To overcome this problem and to improve the code, instead we take the following codewords for $N=3$ photons per block (with $N^{2}=9$ as the total number of photons):
\begin{equation}
\begin{aligned}
 |\bar{0}\rangle=|\widetilde{0}\rangle|\widetilde{0}\rangle|\widetilde{0}\rangle,\\
 |\bar{1}\rangle=|\widetilde{1}\rangle|\widetilde{1}\rangle|\widetilde{1}\rangle,
\end{aligned}
\end{equation}
which are now composed of three blocks for a total number of six modes. To verify that this code corrects all losses up to two photons, we can  calculate the action of AD on the logical qubit. Due to symmetry reasons, it is sufficient to calculate the action of only certain
error operators on the codewords, because all other corrupted codewords with at most two lost photons can be obtained by permutations of the blocks. Therefore, if the KL conditions are fulfilled for the 
following error operators, then they are also satisfied by the block-permuted corrupted states. The relevant error operators are
\begin{widetext}

\begin{align*}
 A_{1}\otimes A_{0}\otimes A_{0}\otimes A_{0}\otimes A_{0}\otimes A_{0}|\bar{\Psi}\rangle
& =\sqrt{\frac{3}{2}\gamma^{8}(1-\gamma)}(\frac{c_{0}}{\sqrt{2}}(|20\rangle+|02\rangle)|\widetilde{0}\rangle|\widetilde{0}\rangle+c_{1}|11\rangle|\widetilde{1}\rangle|\widetilde{1}\rangle),\\
 A_{0}\otimes A_{1}\otimes A_{0}\otimes A_{0}\otimes A_{0}\otimes A_{0}|\bar{\Psi}\rangle
  & =\sqrt{\frac{3}{2}\gamma^{8}(1-\gamma)}(c_{0}|11\rangle|\widetilde{0}\rangle|\widetilde{0}\rangle+\frac{c_{1}}{\sqrt{2}}(|20\rangle+|02\rangle)|\widetilde{1}\rangle|\widetilde{1}\rangle),\\
   A_{1}\otimes A_{1}\otimes A_{0}\otimes A_{0}\otimes A_{0}\otimes A_{0}|\bar{\Psi}\rangle
  & =\sqrt{\frac{3}{2}\gamma^{7}(1-\gamma)^{2}}(c_{0}|01\rangle|\widetilde{0}\rangle|\widetilde{0}\rangle+c_{1}|10\rangle|\widetilde{1}\rangle|\widetilde{1}\rangle),\\
 A_{2}\otimes A_{0}\otimes A_{0}\otimes A_{0}\otimes A_{0}\otimes A_{0}|\bar{\Psi}\rangle
  & =\frac{\sqrt{3}}{2}\sqrt{\gamma^{7}(1-\gamma)^{2}}(c_{0}|10\rangle|\widetilde{0}\rangle|\widetilde{0}\rangle+c_{1}|01\rangle|\widetilde{1}\rangle|\widetilde{1}\rangle),\\ 
  A_{0}\otimes A_{2}\otimes A_{0}\otimes A_{0}\otimes A_{0}\otimes A_{0}|\bar{\Psi}\rangle
  & =\frac{\sqrt{3}}{2}\sqrt{\gamma^{7}(1-\gamma)^{2}}(c_{0}|10\rangle|\widetilde{0}\rangle|\widetilde{0}\rangle+c_{1}|01\rangle|\widetilde{1}\rangle|\widetilde{1}\rangle),\numberthis \\
   A_{1}\otimes A_{0}\otimes A_{1}\otimes A_{0}\otimes A_{0}\otimes A_{0}|\bar{\Psi}\rangle
  & =\frac{3}{2}\sqrt{\gamma^{7}(1-\gamma)^{2}}(c_{0}\frac{1}{\sqrt{2}}(|20\rangle+|02\rangle)\frac{1}{\sqrt{2}}(|20\rangle+|02\rangle)|\widetilde{0}\rangle+c_{1}|11\rangle|11\rangle|\widetilde{1}\rangle),\\ 
  A_{0}\otimes A_{1}\otimes A_{1}\otimes A_{0}\otimes A_{0}\otimes A_{0}|\bar{\Psi}\rangle
  & =\frac{3}{2}\sqrt{\gamma^{7}(1-\gamma)^{2}}(c_{0}|11\rangle\frac{1}{\sqrt{2}}(|20\rangle+|02\rangle)|\widetilde{0}\rangle+c_{1}\frac{1}{\sqrt{2}}(|20\rangle+|02\rangle)|11\rangle|\widetilde{1}\rangle),\\
 A_{0}\otimes A_{1}\otimes A_{0}\otimes A_{1}\otimes A_{0}\otimes A_{0}|\bar{\Psi}\rangle
  & =\frac{3}{2}\sqrt{\gamma^{7}(1-\gamma)^{2}}(c_{0}|11\rangle|11\rangle|\widetilde{0}\rangle+c_{1}\frac{1}{\sqrt{2}}(|20\rangle+|02\rangle)\frac{1}{\sqrt{2}}(|20\rangle+|02\rangle)|\widetilde{1}\rangle),\\
 A_{1}\otimes A_{0}\otimes A_{0}\otimes A_{1}\otimes A_{0}\otimes A_{0}|\bar{\Psi}\rangle
  & =\frac{3}{2}\sqrt{\gamma^{7}(1-\gamma)^{2}}(c_{0}\frac{1}{\sqrt{2}}(|20\rangle+|02\rangle)|11\rangle|\widetilde{0}\rangle+c_{1}|11\rangle\frac{1}{\sqrt{2}}(|20\rangle+|02\rangle)|\widetilde{1}\rangle).\\ 
  \end{align*}

  \end{widetext}
 One can easily verify that a recovery of the logical qubit is, in principle, possible by again detecting the photon number for each block with additional 
 inter-block parity measurements. It is then also not difficult to see that the KL conditions are fulfilled for these operators, so indeed 
 the corresponding two-photon loss errors can be corrected with this encoding. Note that the code is degenerate, i.e. the effect of some non-identical loss
 errors on the logical qubit is identical. For the loss of three or more photons, the code ceases to be a 
 complete loss code.
 The corresponding worst-case fidelity is
\begin{equation}
\begin{aligned}
 F&=\gamma^{9}+9\gamma^{8}(1-\gamma)+36\gamma^{7}(1-\gamma)^{2}\\
 &\approx 1-84(1-\gamma)^{3}.
 \end{aligned}
\end{equation}

\begin{figure}[t!]
\centering
\includegraphics[width=0.51\textwidth]{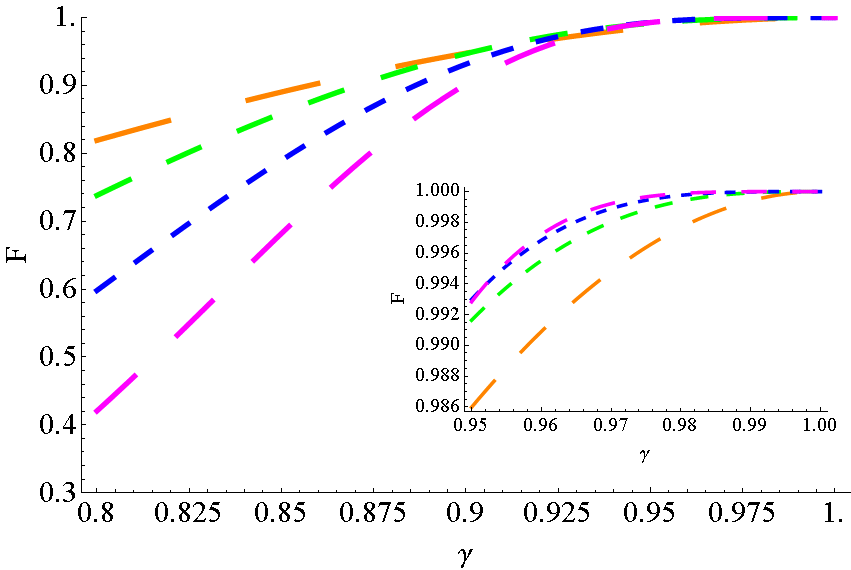}
\caption{Worst-case fidelities for different qubit loss NOON codes as a function of $\gamma$: 
$N^{2}=4$ (orange), $N^{2}=9$ (green), $N^{2}=16$ (blue) and  $N^{2}=25$ (magenta), each correcting $N-1$ photon losses. Notice
the change of ordering with higher-order codes beating the lower-order codes for small losses and the converse for larger losses [see inset].
The small-loss regime $\gamma \in [0.95,1]$ would correspond to a communication channel length of $\sim 1~\text{km}$ (see Section \ref{sec: communication}~).}
\label{fig: qubit-codes}
\end{figure}

\noindent
This is the same result as for the bosonic code in Eq.\eqref{eq: nine}. However, note that in order to promote the encoding from a one-photon-loss
to a two-photon-loss code, in our scheme the maximal photon number per mode only needs to go up from two to three photons (as opposed 
to four versus nine photons in Eq.\eqref{eq:Leung} and Eq.\eqref{eq: nine}, respectively). Similarly, the two-photon-loss code QPC(3,3) requires as many as 18 optical modes
compared to a modest number of six modes in our case.\\
Our procedure can be generalised for arbitrary $N$ (i.e., $N$ photons per block
and $N^{2}$ total number of photons), setting 
\newpage
\begin{equation}
\begin{aligned}
|t_{0}^{2,N}\rangle&=BS[|N0\rangle],\\
|t_{1}^{2,N}\rangle&=BS[|0N\rangle],
\end{aligned}
\end{equation}
 applying the Hadamard-type gate, 
 \begin{equation}
\begin{aligned}
 |\widetilde{0}\rangle&=\frac{1}{\sqrt{2}}\left(|t_{0}^{2,N}\rangle+|t_{1}^{2,N}\rangle\right),\\
 |\widetilde{1}\rangle&=\frac{1}{\sqrt{2}}\left(|t_{0}^{2,N}\rangle-|t_{1}^{2,N}\rangle\right),\\
\end{aligned}
\end{equation}
and finally introducing the $N$-block structure,

\begin{align*}
 |\bar{0}\rangle&=|\widetilde{0}\rangle^{\otimes N}=\left(BS\left[\frac{1}{\sqrt{2}}(|N0\rangle+|0N\rangle)\right]\right)^{\otimes N}\\
 &=\left(\frac{1}{\sqrt{2}^{N-1}}\sum\limits_{j=0}^{N}\sqrt{\binom{N}{2j}}|N-2j,2j\rangle\right)^{\otimes N}, \\
 \numberthis \\
 |\bar{1}\rangle&=|\widetilde{1}\rangle^{\otimes N}=\left(BS\left[\frac{1}{\sqrt{2}}(|N0\rangle-|0N\rangle)\right]\right)^{\otimes N}  \\
  &=\left(\frac{1}{\sqrt{2}^{N-1}}\sum\limits_{j=0}^{N}\sqrt{\binom{N}{2j+1}}|N-2j-1,2j+1\rangle\right)^{\otimes N}.
 \end{align*}
 \noindent
 By construction (for more details, see the next section), this code corrects the loss of up to $N-1$ photons using $N^{2}$ photons. The worst-case fidelities 
 of different qubit codes are compared in Fig.~\ref{fig: qubit-codes} .

\noindent
One interesting feature of our qubit code construction is the interchangeability of the beam splitter transformation, Hadamard operation, and block building. 
For example, consider the $N^{2}=4$ case. In order to produce the codewords, we first apply the symmetric beam splitter transformation 
on $|20\rangle$ and $|02\rangle$, followed by the Hadamard gate, and finally build the blocks. The logical basis codewords obtained in this way are
\begin{equation}
\begin{aligned}
 |\bar{0}\rangle&=\left[\frac{1}{\sqrt{2}}(|20\rangle+|02\rangle)\right]^{\otimes 2}\\
 &=\frac{1}{2}(|2020\rangle+|2002\rangle+|0220\rangle+|0202\rangle),\\
 |\bar{1}\rangle&=\left[\frac{1}{\sqrt{2}}(|20\rangle-|02\rangle\right]^{\otimes 2}\\
 &=\frac{1}{2}(|2020\rangle-|2002\rangle-|0220\rangle+|0202\rangle),
\end{aligned}
\end{equation}
which correspond to the codewords obtained as before up to a beam splitter transformation on each block. The details to verify that this encoding is also a proper code
as well as its extension to qudits can be found in Appendix \ref{sec: alternative noon}~.

\newpage
\section{Generalisation to qudit codes}
\noindent Our method can be directly generalised to logical qudits. Let us again illustrate the idea by a specific example, namely that for a 
qutrit code ($d=3$). Define the states

\begin{align*}
 |t_{0}^{3,2}\rangle&=T[|200\rangle]\\
 &=\frac{1}{3}|200\rangle+\frac{1}{3}|020\rangle+\frac{1}{3}|002\rangle+\frac{\sqrt{2}}{3}|101\rangle\\
 &~~+\frac{\sqrt{2}}{3}|011\rangle+\frac{\sqrt{2}}{3}|110\rangle,\numberthis\\
  |t_{1}^{3,2}\rangle&=T[|020\rangle]\\
 &=\frac{1}{3}|200\rangle+\frac{1}{3}\exp(4\pi i/3)|020\rangle+\frac{1}{3}\exp(-4\pi i/3)|002\rangle\\
 &~~+\frac{\sqrt{2}}{3}\exp(-2\pi i/3)|101\rangle+\frac{\sqrt{2}}{3}|011\rangle\\
 &~~+\frac{\sqrt{2}}{3}\exp(2\pi i/3)|110\rangle,\\
|t_{2}^{3,2}\rangle&=T[|002\rangle]\\
 &=\frac{1}{3}|200\rangle+\frac{1}{3}\exp(-4\pi i/3)|020\rangle+\frac{1}{3}\exp(4\pi i/3)|002\rangle\\
 &~~+\frac{\sqrt{2}}{3}\exp(2\pi i/3)|101\rangle+\frac{\sqrt{2}}{3}|011\rangle\\
 &~~+\frac{\sqrt{2}}{3}\exp(-2\pi i/3)|110\rangle,
  \end{align*}

\noindent  
where $T$ now represents a "tritter" transformation, i.e. a symmetric 3-splitter. 
The encoding works via a qutrit Hadamard-type gate:

\begin{align*}
 |\widetilde{0}\rangle&=\frac{1}{\sqrt{3}}(|t_{0}^{3,2}\rangle+|t_{1}^{3,2}\rangle+|t_{2}^{3,2}\rangle) \numberthis\\
 &=\frac{1}{\sqrt{3}}|200\rangle+\sqrt{\frac{2}{3}}|011\rangle,\\
 |\widetilde{1} \rangle&=\frac{1}{\sqrt{3}}(|t_{0}^{3,2}\rangle+\exp(2\pi i/3)|t_{1}^{3,2}\rangle+\exp(-2\pi i/3)|t_{2}^{3,2}\rangle)\\
 &=\frac{1}{\sqrt{3}}|020\rangle+\sqrt{\frac{2}{3}}|101\rangle,\\
 |\widetilde{2} \rangle&=\frac{1}{\sqrt{3}}(|t_{0}^{3,2}\rangle+\exp(-2\pi i/3)|t_{1}^{3,2}\rangle+\exp(2\pi i/3)|t_{2}^{3,2}\rangle)\\
 &=\frac{1}{\sqrt{3}}|002\rangle+\sqrt{\frac{2}{3}}|110\rangle.\\
\end{align*}

The logical qutrit state is then defined as
\begin{equation}
 \bar{\Psi}\rangle=c_{0}|\widetilde{0}\rangle|\widetilde{0}\rangle+c_{1}|\widetilde{1} \rangle|\widetilde{1} \rangle+ c_{2} |\widetilde{2}  \rangle| \widetilde{2} \rangle.
\end{equation}
\noindent
The states obtained from the logical qutrit after the loss of exactly one photon are:

\begin{align*}
 &A_{1}\otimes A_{0}\otimes A_{0}\otimes A_{0} \otimes A_{0}\otimes A_{0} |\bar{\Psi}\rangle\\
 &=\sqrt{\frac{2}{3}\gamma^{3}(1-\gamma)}(c_{0}|100\rangle|\widetilde{0}\rangle
 + c_{1}|001\rangle|\widetilde{1} \rangle+ c_{2}|010\rangle|\widetilde{2}\rangle),\\
&A_{0}\otimes A_{1}\otimes A_{0}\otimes A_{0} \otimes A_{0}\otimes A_{0} |\bar{\Psi}\rangle\\
&=\sqrt{\frac{2}{3}\gamma^{3}(1-\gamma)}(c_{0}|001\rangle|\widetilde{0}\rangle
 +c_{1} |010\rangle|\widetilde{1} \rangle+c_{2} |100\rangle|\widetilde{2} \rangle),\\
 &A_{0}\otimes A_{0}\otimes A_{1}\otimes A_{0} \otimes A_{0}\otimes A_{0} |\bar{\Psi}\rangle\\
 &=\sqrt{\frac{2}{3}\gamma^{3}(1-\gamma)}(c_{0}|010\rangle|\widetilde{0}\rangle\numberthis 
 +c_{1} |100\rangle|\widetilde{1} \rangle +c_{2}|001\rangle |\widetilde{2} \rangle),\\
 &A_{0}\otimes A_{0}\otimes A_{0}\otimes A_{1} \otimes A_{0}\otimes A_{0} |\bar{\Psi}\rangle\\
 &=\sqrt{\frac{2}{3}\gamma^{3}(1-\gamma)}(c_{0}|\widetilde{0}\rangle|100\rangle
 +c_{1} |\widetilde{1}\rangle|001\rangle+c_{2} |\widetilde{2} \rangle|010\rangle),\\
 &A_{0}\otimes A_{0}\otimes A_{0}\otimes A_{0} \otimes A_{1}\otimes A_{0} |\bar{\Psi}\rangle\\
 &=\sqrt{\frac{2}{3}\gamma^{3}(1-\gamma)}(c_{0}|\widetilde{0}\rangle|001\rangle
 +c_{1}|\widetilde{1}\rangle|010\rangle+c_{2}|\widetilde{2} \rangle|100\rangle),\\
 &A_{0}\otimes A_{0}\otimes A_{0}\otimes A_{0} \otimes A_{0}\otimes A_{1} |\bar{\Psi}\rangle\\
 &=\sqrt{\frac{2}{3}\gamma^{3}(1-\gamma)}(c_{0}|\widetilde{0}\rangle|010\rangle
 + c_{1}|\widetilde{1}\rangle|100\rangle +c_{2}|\widetilde{2}\rangle|001\rangle).\\
 \end{align*}
\noindent
Again, the KL conditions are obviously fulfilled, so the code can correct the loss of up to one single photon.
As before, the error spaces can be discriminated by identifying in which block the photon was lost and by measuring global inter-block observables (while simple 
inter-block parities no longer work). An extension to higher photon numbers and to higher dimensional quantum systems is natural,
\begin{equation}
|t_{0}^{d,N}\rangle=S_{d}[|N00...0\rangle],..., |t_{d-1}^{d,N}\rangle=S_{d}[|000... 0N\rangle].
\end{equation}
Here, $S_{d}$ represents a $d$-splitter, i.e. a symmetric $d$-port device where $d$ is the number of modes. It is the multi-mode generalisation of a symmetric beam splitter
and the tritter as discussed above (thus $S_{2}=BS$ and $S_{3}=T$). As a linear optical device, it is defined by the linear relation between the annihilation operators of the input modes
$a_{i}, i=1,..., d$ and the annihilation operators of the output modes $b_{i}$:
\begin{equation}
b_{i}=\sum\limits_{j=1}^{d}U_{ij}a_{j}.
\end{equation}
Here, the unitary matrix $U$, connecting the input and output modes and ensuring photon number preservation, is given by 
\begin{equation}
 U_{kl}=\frac{1}{\sqrt{d}}\exp\left(i\frac{2\pi kl}{d}\right).
\end{equation}
\noindent
Then we define the following states:
\begin{equation}
|\widetilde{k}\rangle=\frac{1}{\sqrt{d}}\sum\limits_{j=0}^{d-1}\exp(2\pi i kj/d)|t_{j}^{d,N}\rangle,
 \end{equation}
\noindent
for $k=0,..., d-1$. A general logical qudit is then expressed by the $dN$-mode, $N^{2}$-photon state
\begin{equation}
 |\bar{\Psi}\rangle=c_{0}|\widetilde{0}\rangle^{\otimes N}+c_{1}|\widetilde{1}\rangle^{\otimes N}
+...+c_{d-1}|\widetilde{d-1}\rangle^{\otimes N}.
\end{equation}
By construction, this code can correct up to $N-1$ photon losses. The orthogonality of corrupted codewords, required by the KL conditions, 
is easy to check, because the codewords are built blockwise. The non-deformation criterion, however, requires a more rigorous check. Let us first calculate the input
state $|t_{0}^{d,N}\rangle$ for general $N$ and $d$,

\begin{align*}
 &|N00\cdots\rangle= \frac{a_{1}^{\dagger N}}{\sqrt{N!}}|000...\rangle
 \rightarrow S_{d}[|N000...\rangle]\\
 &=\frac{1}{\sqrt{N!}}\sqrt{\frac{1}{d}}^{N}(a_{1}^{\dagger}+a_{2}^{\dagger}+\cdots+a_{d}^{\dagger})^{N}|000\cdots\rangle \\
 &=\frac{1}{\sqrt{N!}}\sqrt{\frac{1}{d}}^{N}\sum\limits_{\vec{k}\in \mathcal{A}} \begin{pmatrix}
                                                                                  N\\
                                                                                  k_{1},k_{2},\cdots, k_{d}
                                                                                 \end{pmatrix}
      a_{1}^{\dagger k_{1}} a_{2}^{\dagger k_{2}}\cdots a_{d}^{\dagger k_{d}}|000\rangle\numberthis \\
&=\frac{1}{\sqrt{N!}}\sqrt{\frac{1}{d}}^{N}\sum\limits_{\vec{k}\in \mathcal{A}} \begin{pmatrix}
                                                                                  N\\
                                                                                  k_{1},k_{2},\cdots, k_{d}
                                                                                 \end{pmatrix}\\
     &\times \sqrt{k_{1}!}\sqrt{k_{2}!}\cdots\sqrt{k_{d}!}|k_{1}, k_{2}, \cdots, k_{d}\rangle\\      
&=\frac{1}{\sqrt{N!}}\sqrt{\frac{1}{d}}^{N}\sum\limits_{\vec{k}\in \mathcal{A}} \frac{N!}{\sqrt{k_{1}!k_{2}!\cdots k_{d}!}}
                                                                                  |k_{1},k_{2},\cdots, k_{d}\rangle.    
                                                                                  \end{align*}
\noindent
In the third line, we used the multinomial theorem, bearing in mind that all the creation operators commute with each other.
Furthermore, we defined the multinomial coefficient, 

\begin{equation}
\begin{pmatrix}
 N\\
 k_{1},k_{2},\cdots, k_{d}
\end{pmatrix}=\frac{N!}{k_{1}!k_{2}!\cdots k_{d}!},
\end{equation}
as the number of arrangements of $N$ objects in which there are $k_{j}$ objects of type j,  $k_{q}$ objects of type q and so on.
We also introduced the set of $d$-dimensional vectors with fixed column sum, i.e. $\mathcal{A}\equiv \{\vec{k}\in \mathbb{N}_{0}^{d}|\sum\limits_{i=1}^{d}k_{i}=N\}$,
to parametrise the set of all $d$-mode Fock states with fixed photon number $N$. Furthermore, we define
$\mathcal{A}^{\prime}\equiv \{\vec{k}\in \mathbb{N}_{0}^{d}|\sum\limits_{i=1}^{d}k_{i}=N~ \text{and}~ k_{1}\geq 1\}$
and $\mathcal{A}^{\prime\prime}\equiv \{\vec{k}\in \mathbb{N}_{0}^{d}|\sum\limits_{i=1}^{d}k_{i}=N-1\}$.\\
We consider the loss of exactly one photon in the first mode, i.e. we apply the operator $A_{1}\otimes A_{0}^{\otimes d-1}$:     

\begin{align*}
 &A_{1}\otimes A_{0}^{\otimes d-1}S_{d}[|N00...\rangle]\\
 &=\frac{1}{\sqrt{N!}}\sqrt{\frac{1}{d}}^{N}\sum\limits_{\vec{k}\in \mathcal{A}} \frac{N!}{\sqrt{k_{1}!}\sqrt{k_{2}!}\cdots\sqrt{k_{d}!}}\\
 &\times A_{1}\otimes A_{0}^{\otimes d-1}|k_{1},k_{2},\cdots, k_{d}\rangle \\  
 &=\frac{1}{\sqrt{N!}}\sqrt{\frac{1}{d}}^{N}\sum\limits_{\vec{k}\in \mathcal{A^{\prime}}} \frac{N!}{\sqrt{k_{1}!}\sqrt{k_{2}!}\cdots\sqrt{k_{d}!}}\\
 &\times\sqrt{\gamma}^{N-1}\sqrt{1-\gamma}\sqrt{k_{1}}|k_{1}-1, k_{2},\cdots, k_{d}\rangle\\  
 &=\frac{1}{\sqrt{N!}}\sqrt{\frac{1}{d}}^{N}\sqrt{\gamma}^{N-1}\sqrt{1-\gamma}\\
 &\times\sum\limits_{\vec{k}\in \mathcal{A^{\prime}}} \frac{N!}{\sqrt{(k_{1}-1)!}\sqrt{k_{2}!}\cdots\sqrt{k_{d}!}}|k_{1}-1, k_{2},\cdots, k_{d}\rangle\\  
 &=\frac{1}{\sqrt{N!}}\sqrt{\frac{1}{d}}^{N}\sqrt{\gamma}^{N-1}\sqrt{1-\gamma}\\
 &\times\sum\limits_{\vec{q}\in \mathcal{A^{\prime \prime}}} \frac{N!}{\sqrt{q_{1}!}\sqrt{q_{2}!}\cdots\sqrt{q_{d}!}}|q_{1}, q_{2},\cdots, q_{d}\rangle\\   
 &=\sqrt{N}\sqrt{\frac{1}{d}}\sqrt{\gamma}^{N-1}\sqrt{1-\gamma}S_{d}(|N-1,0,0,\cdots\rangle)\numberthis                                                                              
\end{align*}

\noindent
For symmetry reasons, the loss of a photon in a different mode acts identically. The same is true for the other input states, i.e. $S_{d}[|0,0,\cdots,N,0,0,\cdots\rangle]$
decays into $S_{d}[|0,0,\cdots,N-1,0,0,\cdots\rangle]$ after losing one photon. Higher losses can be treated by induction. Because the blocks of the basis codewords are exactly 
superpositions of these states, no deformation can take place after photon loss. Together with the orthogonality of corrupted codewords, this proves our qudit encoding to be a quantum 
error correction code.
\section{Physical implementation}
\noindent In order to substantiate the importance of the encodings, we describe a scheme how to generate an arbitrary logical qubit for the simplest
code with just two photons per block ($N=2$). We assume that the states $\frac{1}{\sqrt{2}}(|20\rangle\pm|02\rangle)$ are experimentally accessible
from two single-photon states $|1\rangle\otimes |1\rangle$ with a phase-free and an appropriately phase-inducing, 50:50 beam splitter.
In addition, we need one auxiliary photon in two ancilla modes to produce the following states:
\begin{equation}
\begin{aligned}
 |\psi_{1}\rangle=|0\rangle\frac{1}{\sqrt{2}}(|20\rangle+|02\rangle)|1\rangle,\\
 |\psi_{2}\rangle=|1\rangle\frac{1}{\sqrt{2}}(|20\rangle-|02\rangle)|0\rangle.\\
\end{aligned}
\end{equation}
As pointed out in \cite{sharypov}, by employing an ancilla ion-trap system, the generation of a symmetric entangled state,$\frac{1}{\sqrt{2}}(|\phi_{1}\rangle|\phi_{2}\rangle+|\phi_{2}\rangle|\phi_{1}\rangle),$
is, in principle, possible for arbitrary photonic input states $|\phi_{1}\rangle$ and $|\phi_{2}\rangle$. Applied to $|\psi_{1}\rangle$
and $|\psi_{2}\rangle$, one obtains

\begin{align*}
 &\frac{1}{\sqrt{2}}\left(\frac{1}{\sqrt{2}}(|20\rangle+|02\rangle)\frac{1}{\sqrt{2}}(|20\rangle-|02\rangle)|0110\rangle\right.\numberthis \\
&\left.+\frac{1}{\sqrt{2}}(|20\rangle-|02\rangle)\frac{1}{\sqrt{2}}(|20\rangle+|02\rangle)|1001\rangle\right),
\end{align*}

\noindent
where we already reordered the modes. The next step is to apply a general beam splitter with transmittance $t$, with the coefficients in the desired superposition determined
later, to the first and second pair of the ancilla modes. This leads to

\begin{align*}
&\frac{1}{\sqrt{2}}\left(\frac{1}{\sqrt{2}}(|20\rangle+|02\rangle)\frac{1}{\sqrt{2}}(|20\rangle-|02\rangle)\right.\\
&\times(\sqrt{1-t}|10\rangle-\sqrt{t}|01\rangle)(\sqrt{t}|10\rangle+\sqrt{1-t}|01\rangle)\numberthis \\
&+\frac{1}{\sqrt{2}}(|20\rangle-|02\rangle)\frac{1}{\sqrt{2}}(|20\rangle+|02\rangle) \\
&\left. (\sqrt{1-t}|01\rangle+\sqrt{t}|10\rangle)(-\sqrt{t}|01\rangle+\sqrt{1-t}|10\rangle)\right).
\end{align*}

\noindent
Measuring the photons after the beam splitter and detecting $'1001'$ projects the state onto
\begin{equation}
\begin{aligned}
 &\frac{1-t}{\sqrt{t^{2}+(1-t)^{2}}}\frac{1}{\sqrt{2}}(|20\rangle+|02\rangle)\frac{1}{\sqrt{2}}(|20\rangle-|02\rangle))\\
 &-\frac{t}{\sqrt{t^{2}+(1-t)^{2}}}\frac{1}{\sqrt{2}}(|20\rangle-|02\rangle)\frac{1}{\sqrt{2}}(|20\rangle+|02\rangle)).
\end{aligned}
\end{equation}
Finally, a phase shift of $\pi/2$ on the last mode gives the logical qubit (, i.e applying $\exp\left(\frac{i\pi \hat{n}}{2}\right)$ to it)

\begin{align*}
|\bar{\Psi}\rangle&=\frac{1-t}{\sqrt{t^{2}+(1-t)^{2}}}\frac{1}{\sqrt{2}}(|20\rangle+|02\rangle)\frac{1}{\sqrt{2}}(|20\rangle+|02\rangle))\\
&-\frac{t}{\sqrt{t^{2}+(1-t)^{2}}}\frac{1}{\sqrt{2}}(|20\rangle-|02\rangle)\frac{1}{\sqrt{2}}(|20\rangle-|02\rangle)) \\
&=c_{0}(t)\frac{1}{\sqrt{2}}(|20\rangle+|02\rangle)\frac{1}{\sqrt{2}}(|20\rangle+|02\rangle))\numberthis \label{eq:alternative} \\
&c_{1}(t)\frac{1}{\sqrt{2}}(|20\rangle-|02\rangle)\frac{1}{\sqrt{2}}(|20\rangle-|02\rangle)),
\end{align*}

\noindent
similar to Eq.\eqref{twophotons}. This means that with an appropriated choice of $t$ and a final symmetric beam splitter transformation on the blocks,
any superposition of the logical codewords can be generated. Note that the logical qubit in Eq.\eqref{eq:alternative}
(without the final symmetric beam splitter) corresponds to the four-photon, alternative NOON code qubit [see Section \ref{sec: Qubit codes} and Appendix \ref{sec: alternative noon}~]. 

\section{Application in a one-way communication scheme}
\label{sec: communication}
\begin{figure}[t!]
\centering
\includegraphics[width=0.5\textwidth]{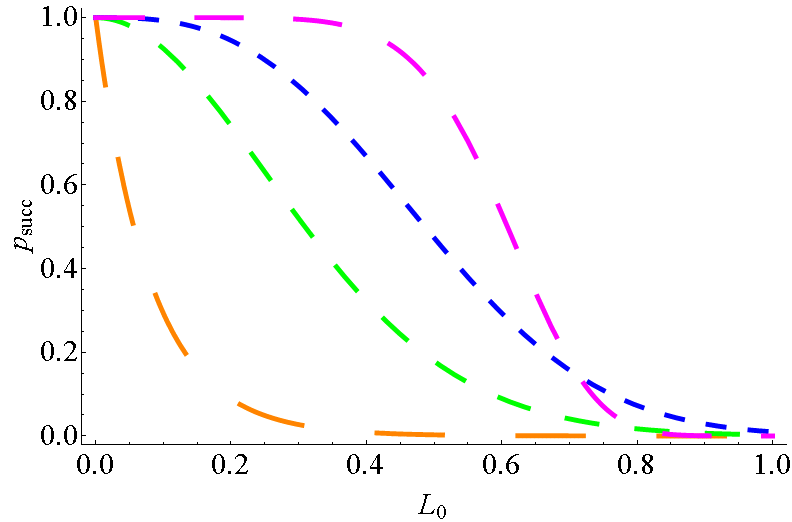}
\caption{Success probabilities for the one-way scheme with different encodings: 
$N^{2}=4$ (orange), $N^{2}=9$ (green), $N^{2}=16$ (blue) and $N^{2}=100$ (magenta) total photons  for $L=1000~ \text{km}$.}
\label{fig: compare}
\end{figure}

\begin{figure}[t!]
  \centering
  \subfigure[$N^{2}=1$]{\includegraphics[width=0.22\textwidth]{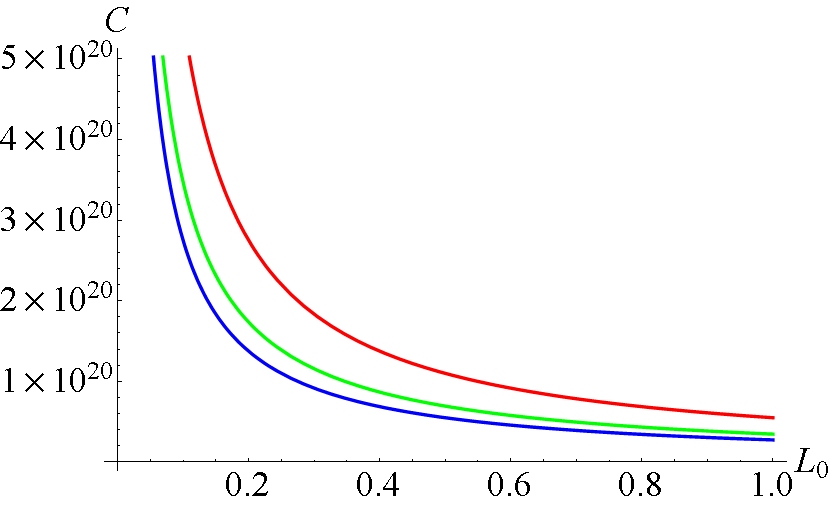}}\quad
  \subfigure[$N^{2}=9$]{\includegraphics[width=0.22\textwidth]{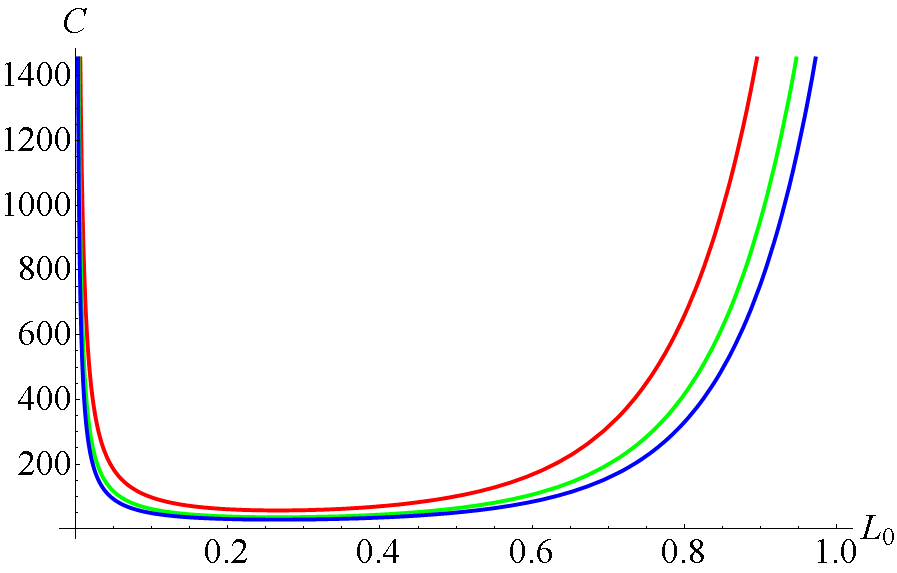}}\quad
  \subfigure[$N^{2}=16$]{\includegraphics[width=0.22\textwidth]{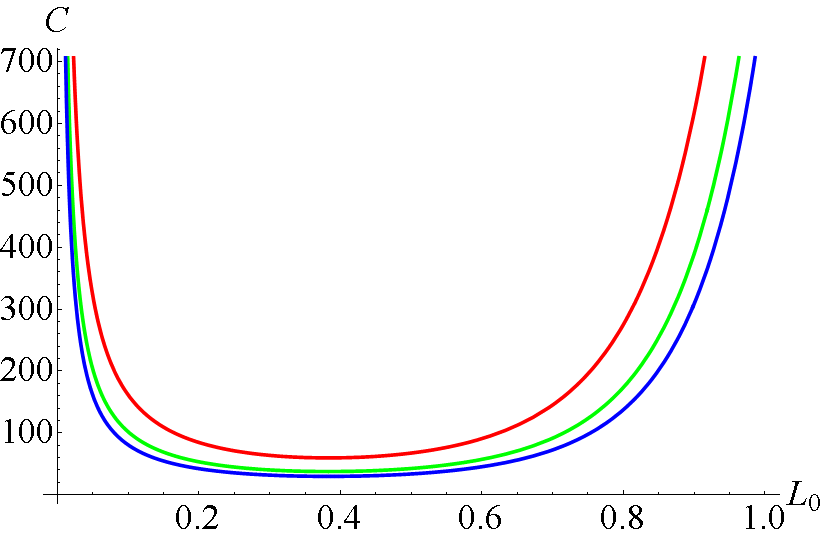}}\quad
  \subfigure[$N^{2}=25$]{\includegraphics[width=0.22\textwidth]{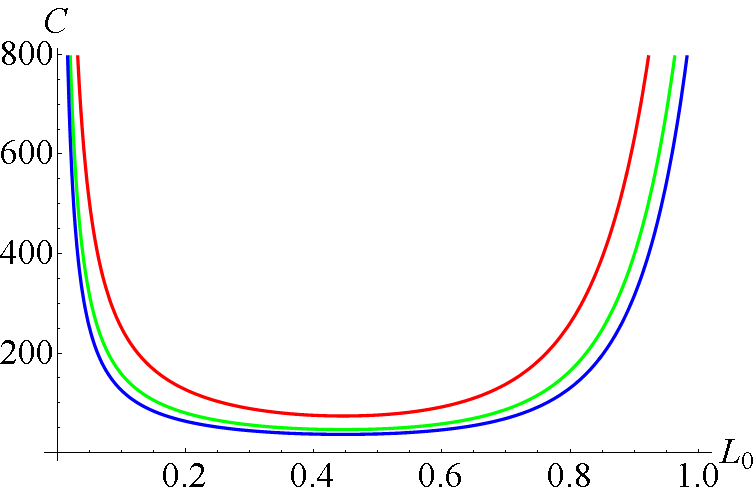}}
  \caption{Cost function for codes with different photon numbers and dimensions for $L=1000~ \text{km}$:
  $d=2$ (red), $d=3$ (green), $d=4$ (blue) (from top to bottom).} 
  \label{fig: cost}
\end{figure}

\noindent In practice, especially for quantum communication, the direct transmission of a photonic state is performed through a noisy quantum channel which leads to an exponential decay of the success rate with the total distance due to photon loss. 
To overcome this problem, besides the standard quantum repeater \cite{Briegel1, Briegel2}, a one-way quantum communication scheme can be applied \cite{lutkenhaus}. Here, an encoded quantum state is sent from a sending station directly through an optical
fibre of length $L_{0}$ to reach the first repeater station while suffering from a moderate amount of photon loss for sufficiently small $L_{0}$.  In each intermediate station, teleportation-based error correction (TEC)\cite{Knill} is performed
before the corrected state is sent to the next repeater station. For logical qubits, TEC is realised by Bell-state preparation
and Bell measurements at the encoded level, which requires encoded Pauli operations as well as encoded Hadamard and CNOT gates.
As pointed out in \cite{Jiang2}, TEC can be generalised to logical qudits using qudit Pauli and SUM gates together with qudit Hadamard gates.\\ 
Based on the results of the last sections, the success probability for one-way communication over a total distance $L$
with repeater spacings $L_{0}$ of an $(N^{2},d)$ encoded qudit is \footnote{The success probability corresponds to a multiple of the fidelity 
calculated in the last sections, $P_{succ}=[F(L_{0})]^{L/L_{0}}$.}  \\
\begin{equation}
 P_{succ}=\left(\sum\limits_{k=0}^{N-1}\binom{N^{2}}{k}\gamma^{N^{2}-k}(1-\gamma)^{k}\right)^{L/L_{0}}. 
\end{equation}
Here, the damping parameter is given by $\gamma=\exp\left(-\frac{L_{0}}{L_{att}}\right)$ with the attenuation length $L_{att}=22~\text{km}$ for telecom fibres
and photons at telecom wavelengths. Note that $P_{succ}$ only depends on $N$ and especially not on $d$. The success probability
for the one-way scheme over a total distance of 1000 km using various codes is shown in Fig.~\ref{fig: compare}  .\\ 
To assess the resources needed in a scheme with our qudit codes, we furthermore define a (spatial) cost function as \cite{lutkenhaus}
\begin{equation}
 C(N,d)=\frac{N^{2}}{P_{succ}\log_{2}(d)L_{0}},
\end{equation}
which depends on the photon number $N$ per block and the dimension of the qudit $d$ \footnote{Compared to \cite{lutkenhaus}, here
we shall only consider the cost for transmitting logical qubits over a total distance $L$, instead of secure classical bits eventually obtained via quantum
communication.}. Fixing the total
photon number, the cost is obviously suppressed by the inverse of the binary logarithm of the qudit dimension (corresponding
to the effective number of encoded logical qubits) such that qudit encodings
make the one-way scheme more efficient. More interesting is the comparison of different qudit encodings with different total photon numbers, as 
shown in Fig.~\ref{fig: cost}~. The plot shows the cost functions of various codes. The cost decreases with $N$ as $P_{succ}$ is increasing
at the same time for a suitably chosen $L_{0}$. 

\noindent
Note that also $N^{2}=1$ can be realised in the so-called multiple-rail qudit encoding, where a single photon occupies
one of $d$ modes, i.e. $|\bar{0}\rangle=|1000..\rangle, |\bar{1}\rangle=|0100..\rangle,..., |\overline{d-1}\rangle=|00...01\rangle$. Since the scaling
of the transmission probability with the loss parameter $\gamma$ only depends on the total photon number (and especially not on the qudit dimension $d$), 
a cost reduction can be achieved already in this case by increasing $d$. However, the multiple-rail encoding is not a quantum error correction code; it is
only a quantum error detection code \cite{N-C} that can detect but not correct loss errors.

\section{Conclusions}
\noindent
We presented a systematic approach for constructing a class of exact quantum error correcting codes for the amplitude-damping channel. Based on
quantum optical NOON state resources, logical qubits can be encoded in a block code consuming a total of $N^{2}$ photons in $N$ blocks. 
These codes are capable of correcting $N-1$ photon losses, which is the same scaling obtainable with existing exact loss codes for the same fixed
total photon number. Nonetheless, only our codes have a total mode number and a maximal photon number per mode that both scale linearly with $N$.\\
All our codes have logical codewords that can be built from NOON states with linear optics. A method for the experimental
generation of the $N^{2}=4$ qubit code including arbitrary logical qubits was also proposed. This method relies on the presence of an ion-trap ancilla system. 
Furthermore, the NOON code approach can be generalised to logical qudits of arbitrary dimension by increasing the mode number per block without
losing the loss robustness, i.e. the fidelity always only depends on the total photon number $N^{2}$ and not on the dimension of the logical qudit.\\
As for an application, this feature is exploited in a one-way communication scheme where general qudit codes turn out to be beneficial in terms of
the spatial resource cost. Limitations of our codes are that there is no simple and efficient method known for the experimental generation of qubit codes
with higher loss resistance and for that of arbitrary qudit codes (including arbitrary logical quantum states). 
This is, however, necessary for the presented one-way scheme, because for achieving a useful success probability at moderate intermediate  distances
$L_{0} \thicksim 1~ \text{km}$,  $N^{2}=\mathcal{O}(100)$ will be required.
In addition, the proposed QND-type measurement for syndrome identification and the corresponding recovery operation, possibly implemented via encoded
qudit quantum teleportation, are experimentally hard to achieve. 

\section{Acknowledgements}
\noindent We acknowledge support from Q.com (BMBF) and Hipercom (ERA-NET CHISTERA).

\bibliographystyle{apsrev4-1}

\bibliography{NOONcode.bib}

\clearpage
\newpage
\onecolumngrid
\appendix
\section{Inefficiency of the N=3 two-block code}
\label{sec: inefficiency}
By calculating the corrupted codewords, the violation of the KL conditions becomes manifest:

\begin{align*}
 A_{0}\otimes A_{0}\otimes A_{0}\otimes A_{0} |\bar{\Psi}\rangle&=\sqrt{\gamma^{6}}|\bar{\Psi}\rangle,\\
 A_{1}\otimes A_{0}\otimes A_{0}\otimes A_{0} |\bar{\Psi}\rangle
 &=\sqrt{\frac{3}{2}\gamma^{5}(1-\gamma)}\left(c_{0}\frac{1}{\sqrt{2}}(|20\rangle+|02\rangle)|\widetilde{0}\rangle+c_{1}|11\rangle|\widetilde{1}\rangle\right),\\
A_{0}\otimes A_{1}\otimes A_{0}\otimes A_{0} |\bar{\Psi}\rangle
 &=\sqrt{\frac{3}{2}\gamma^{5}(1-\gamma)}\left(c_{0}|11\rangle|\widetilde{0}\rangle+c_{1}\frac{1}{\sqrt{2}}(|20\rangle+|02\rangle)|\widetilde{1}\rangle\right),\\
 A_{0}\otimes A_{0}\otimes A_{1}\otimes A_{0} |\bar{\Psi}\rangle 
 &=\sqrt{\frac{3}{2}\gamma^{5}(1-\gamma)}\left(c_{0}|\widetilde{0}\rangle\frac{1}{\sqrt{2}}(|20\rangle+|02\rangle)+c_{1}|\widetilde{1}\rangle|11\rangle\right),\\
 A_{0}\otimes A_{0}\otimes A_{0}\otimes A_{1} |\bar{\Psi}\rangle
 &=\sqrt{\frac{3}{2}\gamma^{5}(1-\gamma)}\left(c_{0}|\widetilde{0}\rangle|11\rangle+c_{1}|\widetilde{1}\rangle\frac{1}{\sqrt{2}}(|20\rangle+|02\rangle)\right),\\
 A_{1}\otimes A_{1}\otimes A_{0}\otimes A_{0} |\bar{\Psi}\rangle
 &=\sqrt{\frac{3}{2}\gamma^{4}(1-\gamma)^{2}}\left(c_{0}|01\rangle|\widetilde{0}\rangle+c_{1}|10\rangle|\widetilde{1}\rangle\right),\\
 A_{0}\otimes A_{0}\otimes A_{1}\otimes A_{1} |\bar{\Psi}\rangle
 &=\sqrt{\frac{3}{2}\gamma^{4}(1-\gamma)^{2}}\left(c_{0}|\widetilde{0}\rangle|01\rangle+c_{1}|\widetilde{1}\rangle|10\rangle\right),\\
 A_{2}\otimes A_{0}\otimes A_{0}\otimes A_{0} |\bar{\Psi}\rangle
 &=\frac{\sqrt{3}}{2}\sqrt{\gamma^{4}(1-\gamma)^{2}}\left(c_{0}|10\rangle|\widetilde{0}\rangle+c_{1}|01\rangle|\widetilde{1}\rangle\right),\\
 A_{0}\otimes A_{2}\otimes A_{0}\otimes A_{0} |\bar{\Psi}\rangle
 &=\frac{\sqrt{3}}{2}\sqrt{\gamma^{4}(1-\gamma)^{2}}\left(c_{0}|10\rangle|\widetilde{0}\rangle+c_{1}|01\rangle|\widetilde{1}\rangle\right),\\
 A_{0}\otimes A_{0}\otimes A_{2}\otimes A_{0} |\bar{\Psi}\rangle
 &=\frac{\sqrt{3}}{2}\sqrt{\gamma^{4}(1-\gamma)^{2}}\left(c_{0}|\widetilde{0}\rangle|10\rangle+c_{1}|\widetilde{1}\rangle|01\rangle\right),\numberthis \\
 A_{0}\otimes A_{0}\otimes A_{0}\otimes A_{2} |\bar{\Psi}\rangle
 &=\frac{\sqrt{3}}{2}\sqrt{\gamma^{4}(1-\gamma)^{2}}\left(c_{0}|\widetilde{0}\rangle|10\rangle+c_{1}|\widetilde{1}\rangle|01\rangle\right),\\  
 A_{1}\otimes A_{0}\otimes A_{1}\otimes A_{0} |\bar{\Psi}\rangle
 &=\frac{3}{2}\sqrt{\gamma^{4}(1-\gamma)^{2}}\left(c_{0}\frac{1}{\sqrt{2}}(|20\rangle+|02\rangle)\frac{1}{\sqrt{2}}(|20\rangle+|02\rangle))+c_{1}|1111\rangle\right),\\ 
 A_{0}\otimes A_{1}\otimes A_{0}\otimes A_{1} |\bar{\Psi}\rangle
 &=\frac{3}{2}\sqrt{\gamma^{4}(1-\gamma)^{2}}\left(c_{1}\frac{1}{\sqrt{2}}(|20\rangle+|02\rangle)\frac{1}{\sqrt{2}}(|20\rangle+|02\rangle)+c_{0}|1111\rangle\right),\\ 
 A_{0}\otimes A_{1}\otimes A_{1}\otimes A_{0} |\bar{\Psi}\rangle
 &=\frac{3}{2}\sqrt{\gamma^{4}(1-\gamma)^{2}}\left(c_{0}|11\rangle\frac{1}{\sqrt{2}}(|20\rangle+|02\rangle)+c_{1}\frac{1}{\sqrt{2}}(|20\rangle+|02\rangle)|11\rangle\right),\\ 
 A_{1}\otimes A_{0}\otimes A_{0}\otimes A_{1} |\bar{\Psi}\rangle 
 &=\frac{3}{2}\sqrt{\gamma^{4}(1-\gamma)^{2}}\left(c_{1}|11\rangle\frac{1}{\sqrt{2}}(|20\rangle+|02\rangle)+c_{0}\frac{1}{\sqrt{2}}(|20\rangle+|02\rangle)|11\rangle\right).\\
\end{align*}
Note that, for example, the corrupted logical basis states in the last two lines are, in general, not orthogonal such that a recovery is not possible. This means that, besides all one-photon losses,
only certain two-photon-loss errors are correctable which gives a worst-case fidelity of $F\approx 1-9(1-\gamma)^{2}$. This result is still worse compared to our $N=2$ four-photon, two-block code.
\newpage

\section{Alternative NOON code construction}
\label{sec: alternative noon}
\subsection{Qubit Codes}
\noindent
The action of the AD channel on the codewords  $|\bar{0}\rangle=\left[\frac{1}{\sqrt{2}}(|20\rangle+|02\rangle)\right]^{\otimes 2}
 =\frac{1}{2}(|2020\rangle+|2002\rangle+|0220\rangle+|0202\rangle)$
 and  $|\bar{1}\rangle=\left[\frac{1}{\sqrt{2}}(|20\rangle-|02\rangle\right]^{\otimes 2}
 =\frac{1}{2}(|2020\rangle-|2002\rangle-|0220\rangle+|0202\rangle)$ is

\begin{align*}
 A_{0}\otimes A_{0}\otimes A_{0}\otimes A_{0}|\bar{0}\rangle&=\sqrt{\gamma^{4}}|\bar{0}\rangle,\\
 A_{0}\otimes A_{0}\otimes A_{0}\otimes A_{0}|\bar{1}\rangle&=\sqrt{\gamma^{4}}|\bar{1}\rangle,\\
 A_{1}\otimes A_{0}\otimes A_{0}\otimes A_{0}|\bar{0}\rangle&=\sqrt{\gamma^{3}(1-\gamma)}\frac{1}{\sqrt{2}}(|1020\rangle+|1002\rangle),\\
 A_{1}\otimes A_{0}\otimes A_{0}\otimes A_{0}|\bar{1}\rangle&=\sqrt{\gamma^{3}(1-\gamma)}\frac{1}{\sqrt{2}}(|1020\rangle-|1002\rangle),\\
 A_{0}\otimes A_{1}\otimes A_{0}\otimes A_{0}|\bar{0}\rangle&=\sqrt{\gamma^{3}(1-\gamma)}\frac{1}{\sqrt{2}}(|0120\rangle+|0102\rangle),\\
 A_{0}\otimes A_{1}\otimes A_{0}\otimes A_{0}|\bar{1}\rangle&=\sqrt{\gamma^{3}(1-\gamma)}\frac{1}{\sqrt{2}}(-|0120\rangle+|0102\rangle), \numberthis\\
 A_{0}\otimes A_{0}\otimes A_{1}\otimes A_{0}|\bar{0}\rangle&=\sqrt{\gamma^{3}(1-\gamma)}\frac{1}{\sqrt{2}}(|0210\rangle+|2010\rangle),\\
 A_{0}\otimes A_{0}\otimes A_{1}\otimes A_{0}|\bar{1}\rangle&=\sqrt{\gamma^{3}(1-\gamma)}\frac{1}{\sqrt{2}}(-|0210\rangle+|2010\rangle),\\
 A_{0}\otimes A_{0}\otimes A_{0}\otimes A_{1}|\bar{0}\rangle&=\sqrt{\gamma^{3}(1-\gamma)}\frac{1}{\sqrt{2}}(|2001\rangle+|0201\rangle),\\
 A_{0}\otimes A_{0}\otimes A_{0}\otimes A_{1}|\bar{1}\rangle&=\sqrt{\gamma^{3}(1-\gamma)}\frac{1}{\sqrt{2}}(-|2001\rangle+|0201\rangle).\\
\end{align*}

\noindent
Obviously, this also defines a quantum error correcting code which can correct the loss of one photon. Since the fidelity only depends on the 
photon number in the codewords, one obtains the same result as for the other $N^{2}=4$ code.\\ 
Similar to the code construction presented in the main text, the extension to codes with higher loss protection works by building blocks
of NOON states with higher photon number. For a total photon number $N^{2}=9$, the logical basis states read
\begin{equation}
\begin{aligned}
 |\bar{0}\rangle&=\left(\frac{1}{\sqrt{2}}(|30\rangle+|03\rangle)\right)^{\otimes 3}\\
 &=\frac{1}{2\sqrt{2}}(|303030\rangle+|303003\rangle+|300330\rangle+|300303\rangle+|033030\rangle+|033003\rangle+|030330\rangle+|030303\rangle),\\
 |\bar{1}\rangle&=\left(\frac{1}{\sqrt{2}}(|30\rangle-|03\rangle)\right)^{\otimes 3}\\
 &=\frac{1}{2\sqrt{2}}(|303030\rangle-|303003\rangle-|300330\rangle+|300303\rangle-|033030\rangle+|033003\rangle+|030330\rangle-|030303\rangle).
\end{aligned}
 \end{equation}
It is not difficult to show that this encoding also represents a quantum error correction code, this time capable of correcting up to two-photon losses. \\
In general,
\begin{equation}
\begin{aligned}
 |\bar{0}\rangle&=\left(\frac{1}{\sqrt{2}}(|N0\rangle+|0N\rangle)\right)^{\otimes N},\\
 |\bar{1}\rangle&=\left(\frac{1}{\sqrt{2}}(|N0\rangle-|0N\rangle)\right)^{\otimes N},
\end{aligned}
\end{equation}
defines a quantum code correcting $N-1$ photon losses using $N^{2}$ total photons.

\subsection{Qudit codes}
\noindent 
The idea for the qubit code construction can be directly generalised to arbitrary qudit codes. Consider $d=3$ and $N=2$ and the qutrit Hadamard transformation $H_{3}$. Then we choose
\begin{equation}
 \begin{aligned}
  |\widetilde{0}\rangle&=H_{3}(|200\rangle)=\frac{1}{\sqrt{3}}(|200\rangle+|020\rangle+|002\rangle),\\
  |\widetilde{1}\rangle&=H_{3}(|020\rangle)=\frac{1}{\sqrt{3}}(|200\rangle+\exp\left(\frac{2\pi i}{3}\right)|020\rangle+\exp\left(\frac{4\pi i}{3}\right)|002\rangle),\\
  |\widetilde{2}\rangle&=H_{3}(|002\rangle)=\frac{1}{\sqrt{3}}(|200\rangle+\exp\left(\frac{4\pi i}{3}\right)|020\rangle+\exp\left(\frac{8\pi i}{3}\right)|002\rangle),\\
 \end{aligned}
 \end{equation}
and build the blocks to construct the basis codewords,
\begin{equation}
\begin{aligned}
 |\bar{0}\rangle&=|\widetilde{0}\rangle|\widetilde{0}\rangle,\\
 |\bar{1}\rangle&=|\widetilde{1}\rangle|\widetilde{1}\rangle,\\
 |\bar{2}\rangle&=|\widetilde{2}\rangle|\widetilde{2}\rangle.\\
\end{aligned}
\end{equation}
It is easy to check that this is a qutrit quantum error correction code, because the loss of a single photon on an individual block gives
\begin{equation}
\begin{aligned}
 A_{1}\otimes A_{0}\otimes A_{0}|\widetilde{0}\rangle&=A_{1}\otimes A_{0}\otimes A_{0}|\widetilde{1}\rangle,
 =A_{1}\otimes A_{0}\otimes A_{0}|\widetilde{2}\rangle=\frac{1}{\sqrt{3}}\sqrt{\gamma^{3}(1-\gamma)}|100\rangle,\\
  A_{0}\otimes A_{1}\otimes A_{0}|\widetilde{0}\rangle&=\frac{1}{\sqrt{3}}\sqrt{\gamma^{3}(1-\gamma)}|010\rangle,\\
  A_{0}\otimes A_{1}\otimes A_{0}|\widetilde{1}\rangle&=\frac{1}{\sqrt{3}}\sqrt{\gamma^{3}(1-\gamma)}\exp\left(\frac{2\pi i}{3}\right)|010\rangle,\\
  A_{0}\otimes A_{1}\otimes A_{0}|\widetilde{2}\rangle&=\frac{1}{\sqrt{3}}\sqrt{\gamma^{3}(1-\gamma)}\exp\left(\frac{4\pi i}{3}\right)|010\rangle,\\
  A_{0}\otimes A_{0}\otimes A_{1}|\widetilde{0}\rangle&=\frac{1}{\sqrt{3}}\sqrt{\gamma^{3}(1-\gamma)}|001\rangle,\\
  A_{0}\otimes A_{0}\otimes A_{1}|\widetilde{1}\rangle&=\frac{1}{\sqrt{3}}\sqrt{\gamma^{3}(1-\gamma)}\exp\left(\frac{4\pi i}{3}\right)|001\rangle,\\
  A_{0}\otimes A_{0}\otimes A_{1}|\widetilde{2}\rangle&=\frac{1}{\sqrt{3}}\sqrt{\gamma^{3}(1-\gamma)}\exp\left(\frac{8\pi i}{3}\right)|001\rangle,\\
  \end{aligned}
  \end{equation}
which proves the non-deformation of corrupted codewords. The orthogonality is ensured by the block structure.\\
To construct a general qudit code, we set
\begin{equation}
\begin{aligned}
 |\bar{0}\rangle_{N,d}&=[H_{d}(|N000...\rangle)]^{\otimes N},\\
 |\bar{1}\rangle_{N,d}&=[H_{d}(|0N00...\rangle)]^{\otimes N},\\
 &\vdots\\
 |\overline{d-1}\rangle_{N,d}&=[H_{d}(|000...0N\rangle)]^{\otimes N},\\
\end{aligned}
\end{equation}
which is again a qudit code correcting $N-1$ photon losses using $N^{2}$ total photons.
\end{document}